\documentclass[preprint,12pt]{elsarticle}
\pdfoutput=1

\usepackage{graphicx}

\usepackage{rotating}

\usepackage{amssymb}

\usepackage{url}

\newcommand{\params}{\mathcal{P}}

\newcommand{\tanom}{\phi}  
\newcommand{\manom}{{\cal M}}  
\newcommand{\eanom}{E}  

\newcommand{\oper}{\tau}  
\newcommand{\ecc}{e}  
\newcommand{\argp}{\omega} 
\newcommand{\mae}{\manom_0}  

\newcommand{\vsys}{\gamma}  

\newcommand{\sjit}{s}

\newcommand{\oparams}{\mathcal{O}}  
\newcommand{\rvnonlin}{\theta}  

\newcommand{\prior}{\pi}
\newcommand{\post}{p}

\def\hspace{{\cal H}}

\newcommand{\like}{\mathcal{L}}

\newcommand{\be}{\begin{equation}}
\newcommand{\ee}{\end{equation}}
\newcommand{\ceqn}[1]{equation~(\ref{#1})}
\newcommand{\ceq}[1]{(\ref{#1})}

\begin{document}

\begin{frontmatter}



\title{Bayesian Methods for Analysis and Adaptive Scheduling of
Exoplanet Observations}


\author[CU]{Thomas J. Loredo}
\author[Duke]{James O. Berger}
\author[CU]{David F. Chernoff}
\author[Duke]{Merlise A. Clyde}
\author[KC]{Bin Liu}

\address[CU]{Department of Astronomy, Cornell University}
\address[Duke]{Department of Statistical Science, Duke University}
\address[KC]{Kuang-Chi Institute of Advanced Technology}

\begin{abstract}
We describe work in progress by a collaboration of astronomers and
statisticians developing a suite of Bayesian data analysis tools for
extrasolar planet (exoplanet) detection,
planetary orbit estimation, and adaptive scheduling of observations.  Our work
addresses analysis of stellar reflex motion data, where a planet is detected
by observing the ``wobble'' of its host star as it responds to the
gravitational tug of the orbiting planet.  Newtonian mechanics specifies an
analytical model for the resulting time series, but it is strongly nonlinear,
yielding complex, multimodal likelihood functions; it is even more complex
when multiple planets are present.  The parameter spaces range in size from
few-dimensional to dozens of dimensions, depending on the number of planets in
the system, and the type of motion measured (line-of-sight velocity, or
position on the sky).  Since orbits are periodic, Bayesian generalizations of
periodogram methods facilitate the analysis.  This relies on the model
being linearly separable, enabling partial analytical marginalization,
reducing the dimension of the
parameter space.  Subsequent analysis uses adaptive Markov chain Monte Carlo
methods and adaptive importance sampling to perform the integrals required for
both inference (planet detection and orbit measurement), and
information-maximizing sequential design (for adaptive scheduling of
observations).  We present an overview of our current techniques and highlight
directions being explored by ongoing research.
\end{abstract}


\end{frontmatter}

\section{Introduction}
\label{sec:intro}

In the last fifteen years astronomers have discovered about 500 planetary
systems hosted by nearby stars; new systems are announced almost weekly and
the pace of discovery is accelerating.  The data are now
of sufficient quantity and quality that exoplanet science is shifting from
being discovery-oriented to focusing on detailed astrophysical modeling and
analysis of the growing catalog of observations.  Making the most of the data
requires new statistical tools that can fully and accurately account for
diverse sources of uncertainty in the context of complex models.

Planets are small and shine in reflected light, making them much dimmer than
their host stars.  With current technology it is not possible to directly
image exoplanets against the glare of their hosts except in rare cases of
nearby systems with a large planet far from its host star.  The vast majority
of exoplanets are instead detected indirectly.  The most productive technique
to date is the {\em radial velocity (RV) method}, which uses Doppler shifts of
lines in the star's spectrum to measure the line-of-sight velocity component
of the reflex motion of the star in response to the planet's gravitational
pull.  Such ``to-and-fro wobble'' can be measured for planets as small as
a few Earth masses, in close orbits around Sun-like stars.  Space-based
telescopes will soon enable {\em high-precision astrometry} capable of
measuring the side-to-side reflex motion (the {\em Hubble Space Telescope}
Fine Guidance Sensors have already performed such measurements for a few
exceptional systems).  Another indirect technique measures the diminution of
light from the star when a planet {\em transits} in front of the stellar disk.
This requires a fortuitous orbital orientation, but large, space-based
transit surveys monitoring many stars are making the transit method
increasingly productive.\footnote{During the review of this manuscript,
the {\em Kepler} mission announced the discovery of over 1000 candidate
exoplanet systems, observed to have periodic, transit-like events; it
is expected that over 90\% of these may be confirmed as exoplanets by
follow-up RV and other observations.}  Astronomers are using these techniques to
build up a census of the nearby population of extrasolar planetary systems,
both to understand the diversity of such systems, and to look for potentially
habitable ``Earth-like'' planets.

Here we focus on analysis of reflex motion data.
Motivated by the needs of astrometric and RV campaigns being planned in 2000,
Loredo \& Chernoff (2000 (LC00), 2003 (LC03)) described Bayesian approaches
for analyzing observations of stellar reflex motion for detection and
measurement of exoplanets, and for adaptive scheduling of observations (see
also Loredo 2004 (L04) for a pedagogical treatment of adaptive scheduling). 
This work advocated a fully nonlinear Bayesian analysis based on Keplerian
models for the reflex motion (i.e., objects in elliptical orbits around a
center of mass).  LC00 noted that a subset of the orbital parameters appear
linearly and (with appropriate priors) could be marginalized analytically,
reducing dimensionality and simplifying subsequent analysis.  Building on the
work of Jaynes and Bretthorst (see Bretthorst 2001 and references therein),
they described the close relationship between this approach and
periodogram-based methods already in use for planet detection, dubbing the
Bayesian counterpart a {\em Kepler periodogram} (Scargle
independently developed similar ideas at the same time, calling the
resulting tool a {\em Keplerogram}; see Marcy et al.\ (2004)).
They also described how Monte Carlo posterior sampling algorithms could
enable implementation of fully nonlinear Bayesian experimental design
algorithms for adaptive scheduling of exoplanet observations. 
Unfortunately, although numerous exoplanets were discovered by the time of
LC00 and LC03, none of the data were publicly available, and the virtues of
the approach could not be demonstrated with real data.

By 2004 several observing teams were at last releasing high-quality
exoplanet RV data, fueling new interest in Bayesian algorithms. 
Independently of earlier work, Cumming (2004) explored the connections
between conventional periodogram methods and the Bayesian approach (albeit
hampered by an incorrect marginal likelihood calculation); Cumming and
Dragomir (2010) later extended this work, reproducing the
Kepler periodogram of LC00.  Ford (2005, 2008) and Gregory (2005) applied
classic posterior sampling techniques (Metropolis random walk (MRW) and
parallel tempering Markov chain Monte Carlo (MCMC), respectively) to orbit
modeling; Gregory's approach uses a control system to tune the proposal
parameters in a pilot run, and he has recently augmented his algorithm 
to include parameter updates based on genetic algorithms (Gregory 2011). 
Balan \& Lahav (2008) applied an early, approximate adaptive MRW algorithm
to the problem.  Tuomi (2011) used the more rigorous adaptive MRW algorithm
of Haario et al.\ (2001) for posterior sampling; to calculate marginal
likelihoods needed for comparing planet and no-planet models, he used the
algorithm of Chib and Jeliazkov (2001) that estimates marginal likelihoods
from MCMC output.

The work in progress that we describe here builds on our own previous work
and these more recent efforts, aiming to develop algorithms with
significantly improved computational efficiency and less complexity for
users in terms of algorithm tuning.  Our motivation is twofold: (1)~To
enable more thoroughgoing analysis of all exoplanet data, including
thousands of datasets with absent or ambiguous evidence for planets (whose
analysis is important for accurate population-level inferences); (2)~To
enable integration of planet detection and orbit estimation results for
individual systems into other statistical tasks, including population-level
modeling (via hierarchical or multi-level Bayesian modeling), and adaptive
scheduling of future observations.

This paper provides an overview of our work, aiming to introduce
statistician readers to exoplanet data analysis, and astronomer readers
to new algorithms for Bayesian computation.  For simplicity, we
focus on analysis of RV data, although we are applying our methods to
astrometric data as well.  We begin by describing the likelihood function and
priors for Keplerian reflex motion models; these functions underly all
subsequent estimation, detection, prediction, and design calculations.  We
then describe an orbital parameter estimation pipeline, including optimal
scheduling of new observations to improve parameter estimates.  Finally, we
describe planet detection calculations, focusing on a new adaptive importance
sampling algorithm we have developed specifically for calculating marginal
likelihoods needed for Bayesian comparison of exoplanet models.

\section{Likelihood Functions and Priors for Keplerian Reflex Motion Models}
\label{sec:reflex}

To remove the imprint of Earth's rotation and revolution from measurements of
stellar motions, astronomers refer RV and astrometric measurements to the
solar system barycenter, which serves as the origin of an inertial reference
frame (one experiencing no accelerations).  The motion of an exoplanet system
with respect to an inertial frame can be separated into
the motion of the system's center of mass (COM), which moves at a constant
velocity, and the relative motions of the star and planets with respect to the
COM.  To model the relative motions, we assume that the star and planets have
small sizes compared to their separations (allowing us to ignore asphericity
and tidal forces), and that non-gravitational forces are negligible.

For a single-planet system, Newtonian physics predicts that the star and
planet each move in congruent, coplanar, periodic Keplerian orbits, so-named
because the orbits obey a version of Kepler's laws:  (1)~each orbit is an
ellipse with focus at the COM; (2)~the vector from the COM to a body sweeps
out area at a constant rate; (3)~for given masses, orbits of different period
have the square of the period proportional to the cube of the ellipse's
semimajor axis.  For the vast majority of systems we are only able to observe
the star's orbit, but the planet's orbit is a scaled version of the star's,
larger by a factor equal to the star-planet mass ratio.  If the star and
planet masses are known, three additional parameters suffice to describe the
relative motion in the orbital plane:  the {\em semimajor axis}, $a$, and {\em
eccentricity}, $\ecc$, of the ellipse (with the masses, these fix the orbital
period), and an angle specifying the star's position along the orbit at some
reference time (e.g., the mean anomaly parameter, $\mae$, defined below). 
To describe the motion with respect to an arbitrary inertial frame, we also
need the COM velocity, and three additional geometric parameters to specify
the orbit's orientation in space (e.g., via Euler angles).  In practice, the
stellar mass may be estimated to a few percent accuracy from spectral data,
but the planet mass is unknown; in fact, measuring the masses of planets is
a prime scientific goal of exoplanet observations.  The {\em orbital
period}, $\oper$, is then an additional free parameter.

For a multiple planet system, in general there is no simple description of the
motion.  But so long as the star is much more massive than any of the planets,
and if the orbits are not resonant (e.g., with commensurable periods), then the
motion of each planet may be well approximated by a Keplerian orbit about the
COM of the host star and any closer planets (the resulting hierarchy of
COM-referenced coordinates are called Jacobi coordinates).  Such models for
multiple-planet systems are called multi-Kepler models.  Our work considers
only Keplerian and multi-Kepler models, but astronomers can calculate more
accurate multiple-planet orbits using $N$-body algorithms, and such tools are
needed for modeling systems with resonant orbits.

\begin{figure}[t]
\centering
\includegraphics[clip,width=.75\textwidth]{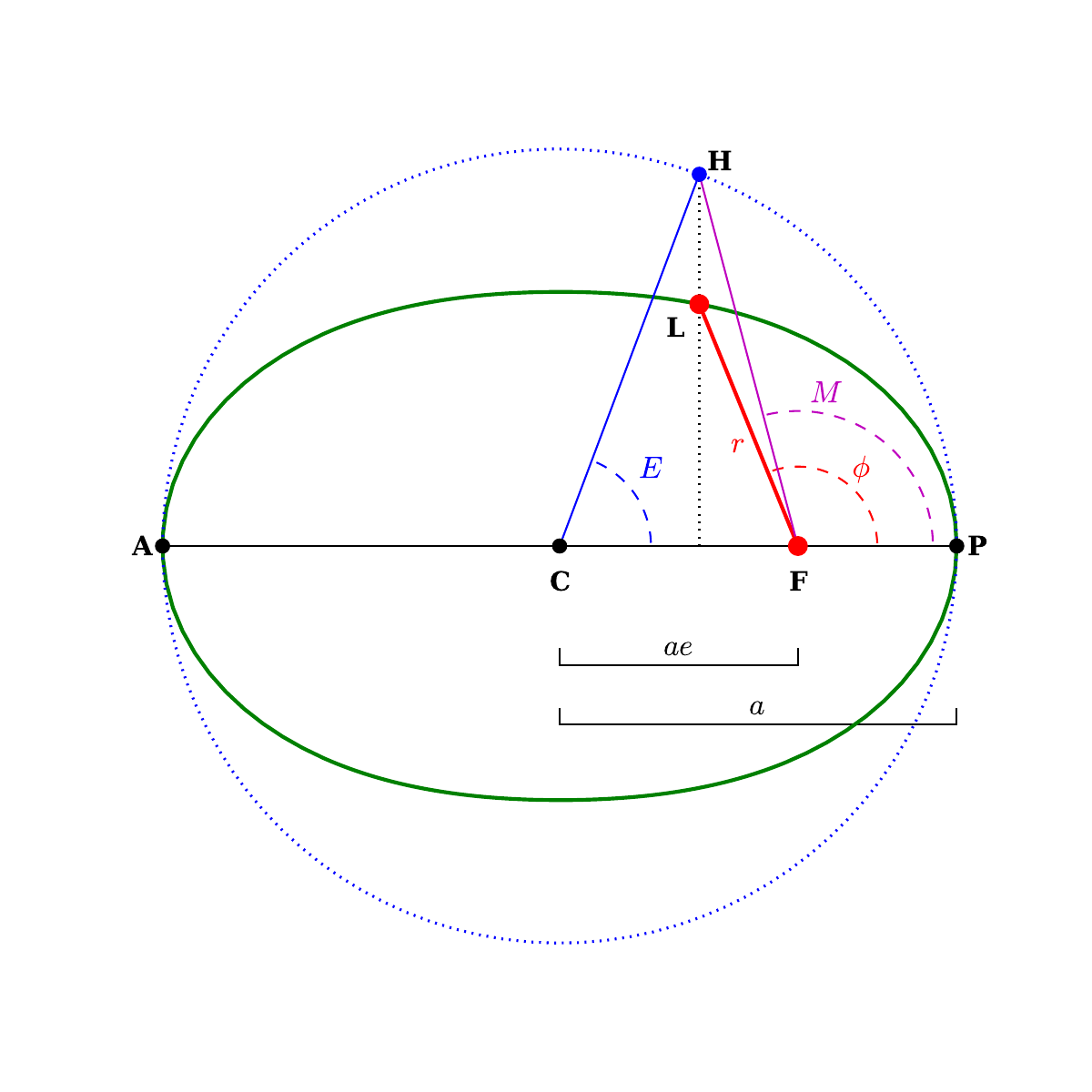}
\caption{Geometry of an elliptical orbit (green curve) with eccentricity
$e=0.6$ and semimajor axis $a$, viewed looking down on the orbital plane.
$F$ is the focus (the center of mass), $P$ is at periapsis, and $C$ is at
the center of the orbit, and of the circumscribing circle of radius $a$
(dotted blue curve).  $L$ represents the location of the object (planet or
star) at a point on its orbit, at a distance $r$ from $F$. $\tanom$ is the
true anomaly for $L$, $M$ is the mean anomaly, and $E$ is the eccentric
anomaly.}
\label{fig:orbplane}
\end{figure}

Returning to the single-planet case, calculating the (vector) velocity of the
star at a given time, $t$, and projecting it along the line of sight, produces
a simple expression for the radial velocity as a function of an angular
coordinate called the {\em true anomaly}, $\tanom(t)$.  The true anomaly
specifies the position of the star on its elliptical orbit via an angle
measured with respect to periapsis (the point of closest approach to the COM),
with the vertex for the angle being the focus of the ellipse (at the COM). 
Figure~\ref{fig:orbplane} shows the geometry.  The predicted radial velocity at
time $t$ is,
\be
v(t) = K \left(\ecc\cos\argp + \cos[\argp + \tanom(t)]\right)
         + \vsys,
\label{v-def}
\ee
where the first term is the Keplerian contribution, and $\vsys$ is a systemic
velocity parameter that combines the projected COM velocity and other 
constant offsets in the velocity measurement.  Here $K$ is the
{\em velocity semi-amplitude}, and $\argp$ is the {\em argument of periapsis},
one of three angular parameters conventionally used to specify the
three-dimensional orientation of the orbit with respect to the frame of
measurement (the other two are nonidentifiable with RV data).

Equation \ceq{v-def}\ appears simple, but significant complication is hidden
in $\tanom(t)$:  it varies in time in a complicated manner, and depends
nonlinearly on three of the orbital parameters.  In fact, no simple, direct
formula for $\tanom(t)$ is known; it may be calculated implicitly using two
other angular coordinates for the orbital position.  First is the {\em mean
anomaly}, $\manom$, another angle with vertex at the focus, but measuring
the planet's position projected (orthogonal to the semi-major axis) on a
circle circumscribing the orbital ellipse.  Figure~\ref{fig:orbplane}
displays this peculiar construction, which makes $\manom(t)$ express
Kepler's second law by varying uniformly in time:
$\manom(t) = 2\pi t/\oper + \mae$,
where $\mae$ is the {\em mean anomaly at epoch} $t=0$, a free parameter.
Second is the {\em eccentric anomaly}, $\eanom$, an angle defined like
$\manom$, but using the center of the ellipse as the vertex; it is also shown
in the figure.  The mean and eccentric anomalies are related via a famous
transcendental equation, Kepler's equation:
\be
\eanom - e\sin \eanom = \manom.
\label{kepler-eqn}
\ee
Once $\eanom$ is found from $\manom$ (at a particular time of interest),
$\tanom$ is given by
\be
\tan{\tanom\over 2} = \left(1+e\over 1-e\right)^{1/2} \tan {\eanom\over 2}.
\label{tan-u-E}
\ee
The time dependence of the true anomaly thus depends on
$\oper$, $\ecc$, and $\mae$.

To estimate the radial velocity of a star at a given epoch, astronomers make
high-resolution measurements of the star's spectrum, with instrumentation
that imposes a number of calibration lines on the spectrum.  Multiple
spectra, spaced closely in time, may be taken to facilitate averaging to
reduce uncertainties.  A complicated joint analysis of all lines (stellar
and calibration) in all spectra yields an estimate of the stellar line
Doppler shift, with uncertainty summarized by a standard deviation; the
uncertainty typically corresponds to a line shift of a tiny fraction of the
width of a spectral pixel, corresponding to velocities of 1--10 m~s$^{-1}$. 
But there is an additional source of velocity uncertainty.  Stellar surfaces
are in constant turbulent motion at high velocities.  Stars are not
spatially resolved in RV observations, so the measured velocity is an
average over the stellar surface (and over the short time span of a single
epoch's observations). Most of the turbulent motion gets averaged away, but
a small, random {\em stellar jitter} component remains, with a standard
deviation, $\sjit$, that varies from star to star and that is typically a
few m~s$^{-1}$ (see Wright 2005 for a compilation of jitter measurements for
apparently planet-free stars).  Further, stars rotate; if starspots are
present, effectively masking part of the stellar surface, a small part of
the rotational motion---``starspot jitter''---will contribute to the RV
measurement (Makarov et al.\ 2009).

An RV data set is thus comprised of a set of $N$ velocity estimates and
associated uncertainties, $\{v_i, \sigma_i\}$, at times $t_i$ (not regularly
spaced), where the measurements may be modeled as the sum of the systemic
and Keplerian stellar velocity, and instrumental and jitter noise.  We
describe the uncertainty in both noise contributions with normal probability
distributions. The resulting likelihood function is,
\begin{equation}
\like(\oparams,\vsys,\sjit) =
   \left[\prod_{i=1}^N \frac{1}{2\pi\sqrt{\sigma_i^2 + \sjit^2}}\right] 
     \exp\left[-\frac{1}{2} \chi^2(\oparams,\vsys,\sjit)\right]
\label{like-def}
\end{equation}
where the quantity in the exponent is the familiar goodness-of-fit statistic,
\be
\chi^2(\oparams,\vsys,\sjit) \equiv 
  \sum_i \frac{[d_i-v(t_i;\oparams,\vsys)]^2}{\sigma_i^2 + \sjit^2}.
\label{chi2-def}
\ee
Here $\oparams$ denotes the five orbital parameters $(K, \tau, \ecc, \argp,
\mae)$ that specify the Keplerian contribution to the velocity curve, $v(t)$.
When we consider multiple-planet models (in Section~4), a separate set of
$\oparams$ parameters appears for each planet, but there is only one $\vsys$
and one $\sjit$ per system.

Conventional analyses of RV data first use a Lomb-Scargle periodogram (LSP;
Black \& Scargle 1982) to identify candidate periods, and then use a
nonlinear minimizer to find a best-fit orbit by minimizing \ceqn{chi2-def}. 
Jitter is not handled as a free parameter ($\chi^2$ minimization cannot be
used to estimate such variance parameters); instead, the instrumental
uncertainties are scaled to make the minimum $\chi^2$ equal to the number of
degrees of freedom (i.e., reduced $\chi^2$ of unity), with the excess
variance attributed to jitter.  Uncertainties are typically estimated either
via the observed Fisher information matrix (whose inverse is asymptotically
a covariance matrix estimate under regularity conditions), or using Wilks's
theorem to select $\chi^2$ contours (which bound confidence regions of
specified asymptotic coverage under regularity conditions).

There are several shortcomings of such conventional
procedures.  The likelihood function is always highly multimodal, and due in
part to the relative sparsity of RV data sets, there are often multiple viable
modes.  Also, the best-fit parameters often lie on or near parameter space
boundaries (especially for $\ecc$).  These features violate the regularity
conditions for use of the Fisher information or Wilks's theorem; the resulting
confidence regions can have grossly incorrect coverage.  Quantities of key
astrophysical interest, including the semimajor axis and the planet mass, are
nonlinear functions of the parameters in the velocity model; propagating the
uncertainties into such lower-dimensional nonlinear spaces is nontrivial.
Uncertainty propagation into predictions (for future observations) and
population-level analyses is similarly challenging.  Also, there is no good
way to incorporate prior information into the analysis, in particular, prior
information on the scale of jitter.  Finally, the LSP is closely related to
least-squares fitting of a sinusoid to the data.  It is not optimal for
identifying candidate orbits of non-negligible eccentricity (in which case the
velocity curve can be very non-sinusoidal).

Bayesian methods can address these shortcomings.  Multiplying the likelihood
function by a prior density, $\prior(\oparams,\vsys,\sjit)$, and normalizing,
produces a posterior density in parameter space,
$\post(\oparams,\vsys,\sjit|D)$, with $D$ denoting the data.  From a set of
samples of $(\oparams,\vsys,\sjit)$ drawn randomly from the posterior, one can
straightforwardly address a wide variety of inference tasks without relying on
unjustifiable asymptotic approximations.  Manipulation of the posterior can
produce a periodogram-like quantity that improves sensitivity to eccentric
orbit signals over the LSP.  Some Bayesian inferences may depend on the choice
of prior when the data are particularly sparse.  But this dependence is
useful:  once data sets from many systems are available, population
properties can be learned via the dependence of system inferences on the
prior; this is called hierarchical or multi-level Bayesian modeling.  Thus for
initial system-by-system inference, we adopt {\em interim priors}, motivated
in part by approximate population properties, and by convenience.  Once enough
systems have been discovered and analyzed, joint analysis of many interim
inferences can be used to learn the system-level prior.

During the 2006 Astrostatistics Program hosted by the Statistical and
Mathematical Sciences Institute (SAMSI), astronomers working on Bayesian
methods for exoplanet science (including our team) settled on ``default''
interim priors for RV model parameters.  They are motivated by basic physical
constraints and approximate population-level knowledge, but attempt to be
otherwise relatively uninformative (e.g., approximately flat or log-flat, but
bounded where necessary to be proper).  See Ford \& Gregory (2007)  for
details.  When it is useful, in our work we sometimes modify default priors,
e.g., using a slightly different $K$ prior to facilitate analytic dimensional
reduction.  Our current completed calculations use an uninformative jitter
prior for comparison with previous work, but we are also performing
calculations with an informative jitter prior (based on the jitter study of
Wright 2005).  This can be important for obtaining realistic Bayes factors
comparing models with different numbers of planets (including no planet);
without an informative jitter prior, evidence for a planet can be masked when
a noninformative prior allows significant residuals to be interpreted as
jitter.

Before developing algorithms to implement the required calculations, it is
important to understand the structure of the likelihood function.  The first
thing to note is that the velocity signal model, \ceqn{v-def}, is a linear
model with respect to two parameters, $K$ and $\vsys$.  Combined with the
Gaussian form of the likelihood function, this implies that the dependence
of the likelihood function with respect to $K$ and $\vsys$ is multivariate
normal, provided other parameters are held fixed.  If the prior for the
linear parameters is approximately flat or normal, any needed integrals
with respect to these parameters may be performed analytically (the result
can be cast in terms of well-known weighted linear least squares solutions).

In fact, a simple reparameterization can reduce the nonlinear dimensionality
further.  Let $A_0 = \vsys$, $A_1 = K\cos\argp$, and $A_2 = -K\sin\argp$.
Then the velocity equation may be written as
\be
v(t) = A_0 + A_1[\ecc + \cos\tanom(t)] + A_2\sin\tanom(t).
\label{amp-model}
\ee
A single-planet model now has three linear parameters, $A\equiv (A_0, A_1,
A_2)$, three nonlinear orbital parameters, $\rvnonlin\equiv(\oper,\ecc,\mae)$,
and the jitter parameter.  If we adopt a flat or normal interim prior on $A$,
we can analytically marginalize over the amplitudes, effecting a significant
dimensional reduction.  (A flat amplitude prior corresponds to a prior that
rises linearly with $K$; if desired, this can be adjusted to another prior
later in the calculation via importance resampling.)

\begin{sidewaysfigure}
\centering
\includegraphics[width=0.52\textwidth]{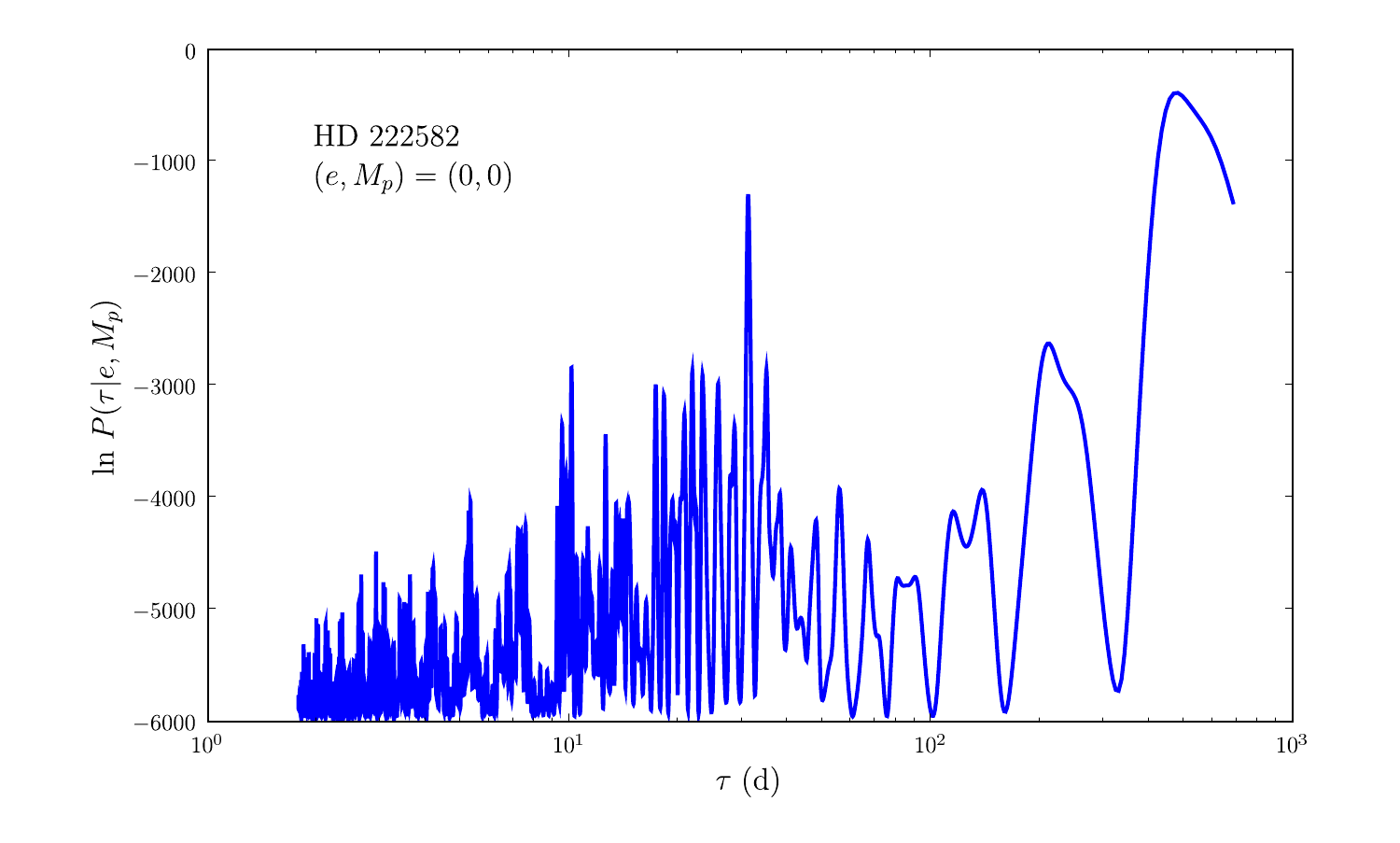}\nobreak~~\nobreak
\includegraphics[width=0.52\textwidth]{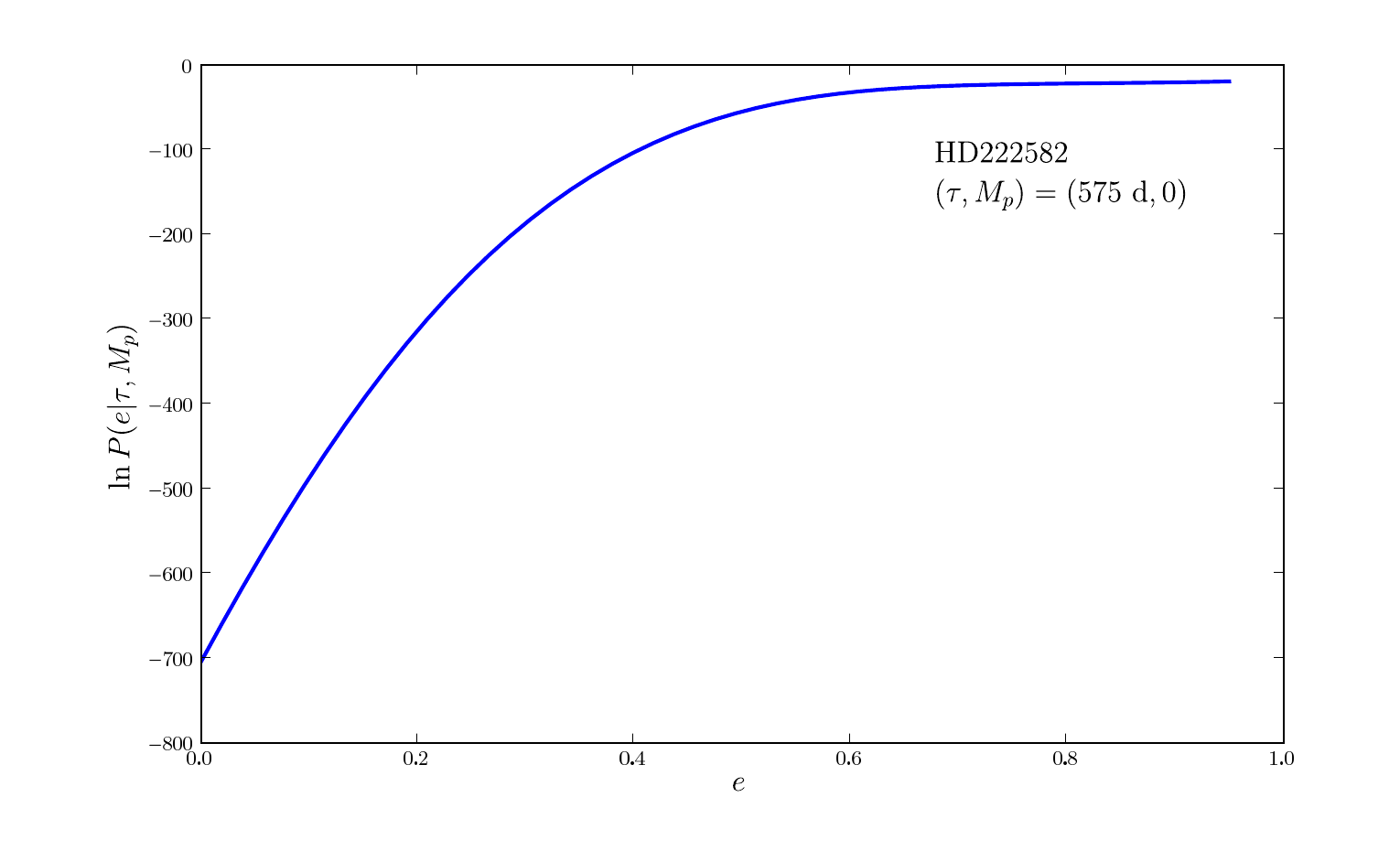}\\
\includegraphics[width=0.5\textwidth]{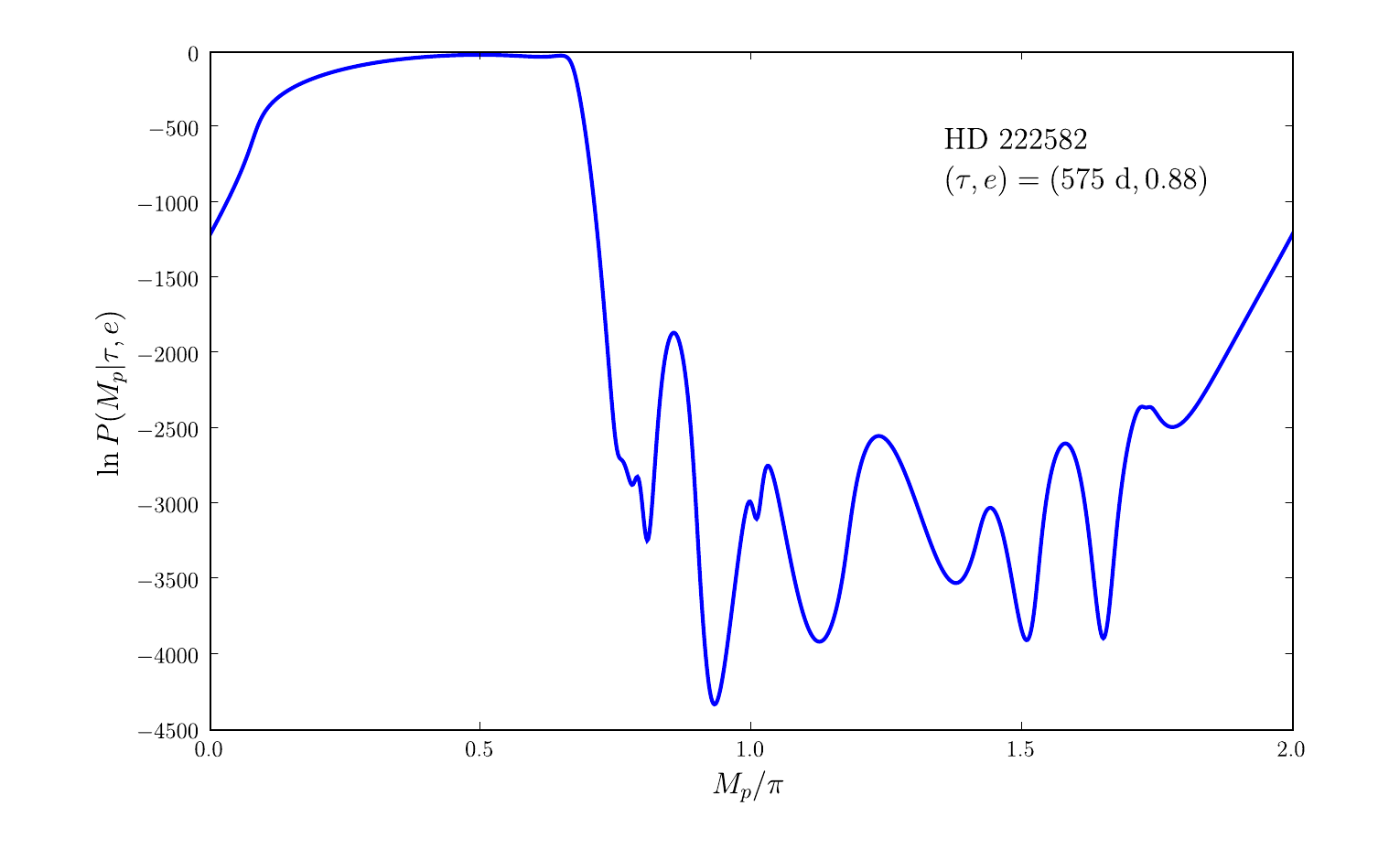}
\caption{\small ``Slices'' of the marginal likelihood function,
$\like_m(\tau,\ecc,\mae)$, for orbital parameters of a single-planet
model for 24 RV observations of HD~222582.}
\label{fig:slices}
\end{sidewaysfigure}

The likelihood function has relatively simple behavior with respect to $A$. To
gain insight into its behavior as a function of the nonlinear parameters,
Figure~\ref{fig:slices} displays slices of the amplitude-marginalized
log-likelihood function along the $\oper$, $e$, and $\mae$ dimensions, based
on 24 RV observations of the single-planet system HD~222582 obtained over a
683~d time span with instrumentation at the Keck observatory (Vogt et al.\
2000; V00). The likelihood function is highly multimodal with respect to
$\oper$, relatively smooth with respect to $\ecc$ (but for many datasets
peaking on or near a boundary), and multimodal with respect to $\mae$.
Computational algorithms must be capable of handling these complications.

Readers familiar with periodogram methods for time series analysis will
recognize the structure exhibited in the $\oper$ slice in
Figure~\ref{fig:slices}.  In fact, for $\ecc=0$ (in which case $\mae$ is
nonidentifiable), the logarithm of the marginal likelihood function for
$\oper$ is proportional to the LSP to a good approximation (this follows from
results in Bretthorst 2001).  Accordingly, we interpret the log marginal
likelihood function for nonlinear parameters, $L_m(\oper,\ecc,\mae)$ (possibly
including jitter when important), as a generalization of the LSP; we dub it a
{\em Keplerogram} or K-gram.  If we numerically integrate the marginal
likelihood over $\ecc$ and $\mae$ (weighted by a flat prior), its logarithm
offers a 1-D summary of the evidence for a planet as a function of candidate
period.  We dub this a {\em Kepler periodogram}.

\section{Orbital Parameter Estimation and Adaptive Scheduling}
\label{sec:orbital}

The pipeline we currently use for orbital parameter estimation takes
advantage of dimensional reduction to facilitate adaptive posterior sampling
for exploring exoplanet models.  We calculate the marginal likelihood
function for the nonlinear parameters, i.e., the exponential of the
Keplerogram, and posterior sampling is done in the reduced-dimension
nonlinear parameter space.  Once samples are available, they can be readily
augmented to the full dimensional space (i.e., with amplitude parameter
samples added) by drawing samples from the conditional multivariate normal
distribution for $A$ given the nonlinear parameters.  If we wish to adjust
the interim prior, importance weights may be assigned to the samples at this
step.  The samples are then used for subsequent inferences.

To initialize a posterior sampler, we fix the jitter to a trial value, and
explore the $(\tau,\ecc,\mae)$ dependence of the marginal posterior via an
adaptive grid in $\tau$ (refining the grid in regions of high posterior
density), supplemented by a crude grid or quadrature in $(\ecc,\mae)$ in
promising $\tau$ regions.  We use this exploration to build a simple,
approximate $(\tau,\ecc,\mae)$ sampler, used for drawing initial values for
rigorous posterior sampling.  For two-planet models, we follow this procedure
for the first planet, and then for each member of a small set of approximate
samples for the first planet's parameters, repeat the procedure for a
single-planet model for the residuals from the candidate model for the first
planet.  This lets us build candidate multi-Kepler models via a tractable
sequence of single-planet fits.  The known multiple-planet systems
were first detected using such a ``hieararchical'' approach, in
the context of the periodogram/$\chi^2$ minimization approach.

\begin{figure}[t]
\centering
\centerline{\includegraphics[clip,width=.7\textwidth]{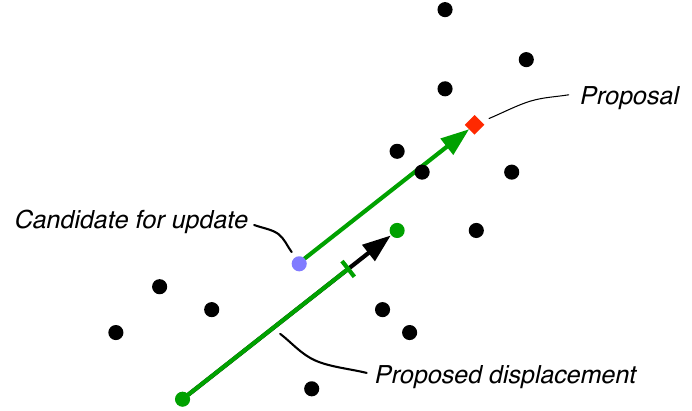}}
\caption{\small Illustration of the differential evolution MCMC algorithm.
Filled circles show 15 points comprising an evolving population in a
two-dimensional parameter space.  Blue circle indicates a point chosen at
random from the 15 for possible updating.  Green circles indicate pair of
points chosen at random from the 14 remaining points; these determine a
proposal displacement, defined as a fraction of the displacement between
them.  Red diamond indicates the proposed update to the blue point, found
using the scaled displacement.  It will be accepted or rejected via the
Metropolis-Hastings rule.}
\label{fig:DEMC}
\end{figure}

We are exploring a few different methods for posterior sampling of the
marginal posterior for the nonlinear parameters.  A particularly promising
and simple method we are using is ter Braak's {\em differential evolution
Markov chain (DEMC)} method (ter Braak 2006).  This is a population-based
adaptive MCMC algorithm that weds ideas from evolutionary computing and
MCMC.  It evolves several chains of samples in the nonlinear parameter
space, all of them targeting the marginal posterior density (in contrast to
the better-known parallel tempering algorithm, where all chains but one
explore annealed versions of the posterior, producing samples that are not
useful in final inferences).  The chains interact at every step in a way
that preserves fair and asymptotically independent sampling in each chain;
but the interaction effectively tunes the size and shape of the proposal
distribution to match features of the target.  Figures~\ref{fig:DEMC}
depicts one DEMC update step.  The algorithm enables exploration of multiple
modes by different chains (partly via hopping between modes for each chain)
provided the initial population includes samples from the important modes.
We attempt to ensure this by building the initial population using the
K-gram.  The aim of the algorithm is to use a small-sized population to
define adaptive Metropolis-Hastings proposals for updating each chain, with
the final output being the samples in all chains.

\begin{figure}
\centering
\includegraphics[width=\textwidth]{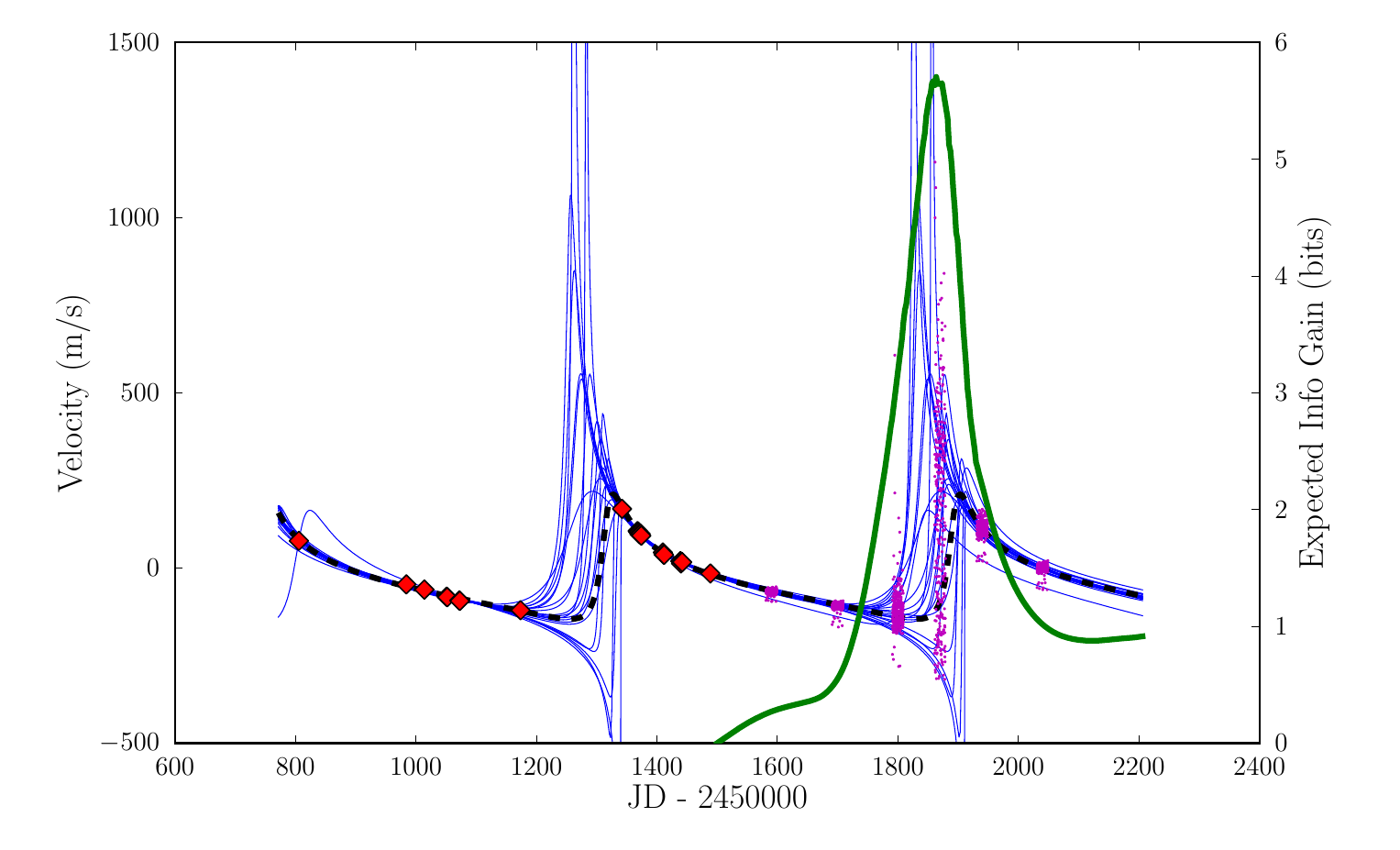}
\includegraphics[width=.9\textwidth]{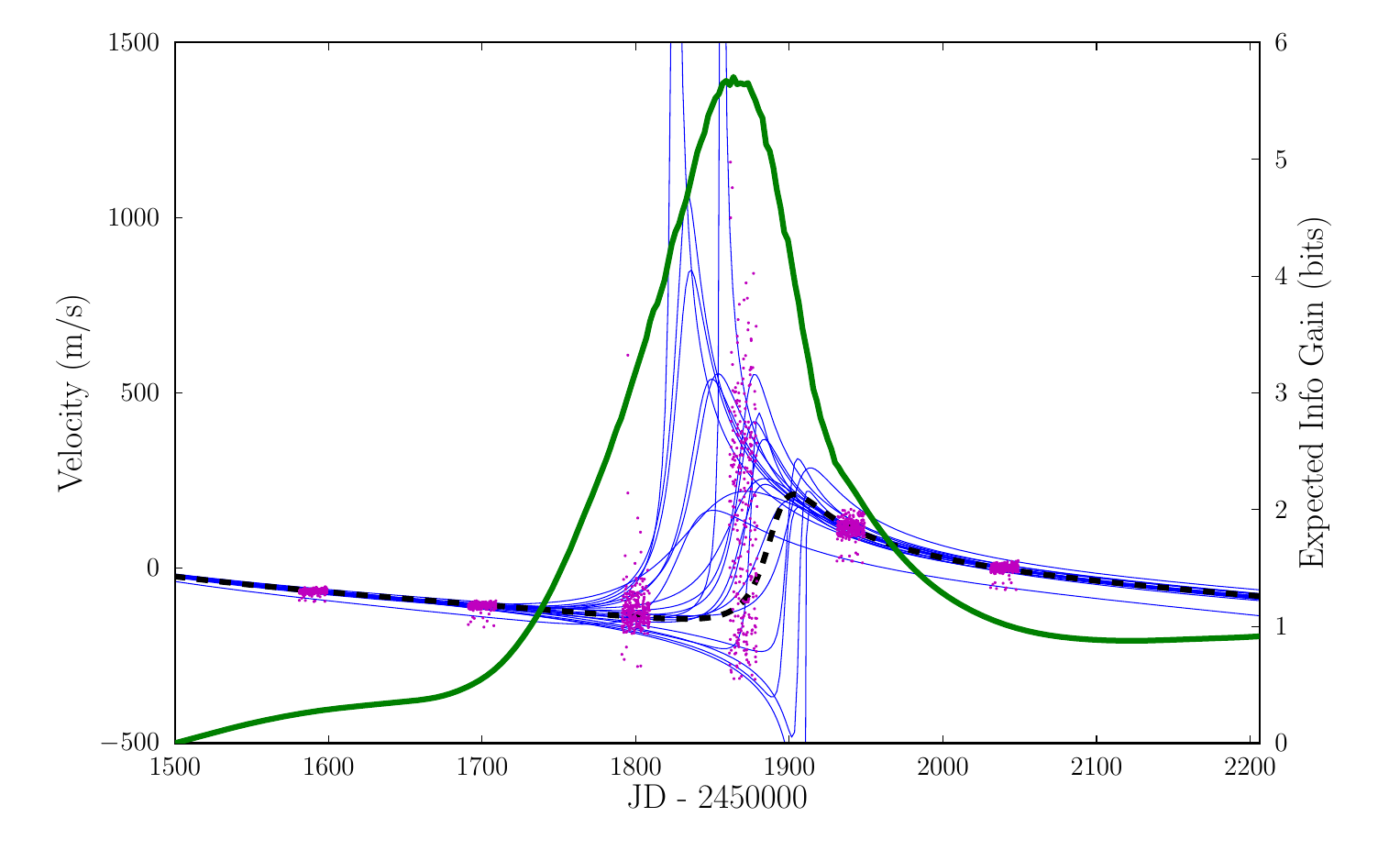}
\caption{\small Results from single-planet modeling of 24 RV observations of
HD~222582 reported in V00 (as revised in B06).  Shown are observations (red
diamonds, several overlapping), best-fit velocity model (thick dashed black
curve), 20 representative models from posterior (thin blue curves), 300
samples from predictive distribution for velocity at six epochs (magenta
points), and relative expected information gain vs.\ future observing time
(green curve, right axis).  Time is barycentric Julian day relative to
2,450,000. Bottom panel expands the later half of the plot.}
\label{fig:HD1a}
\end{figure}

\begin{figure}
\centering
\includegraphics[width=\textwidth]{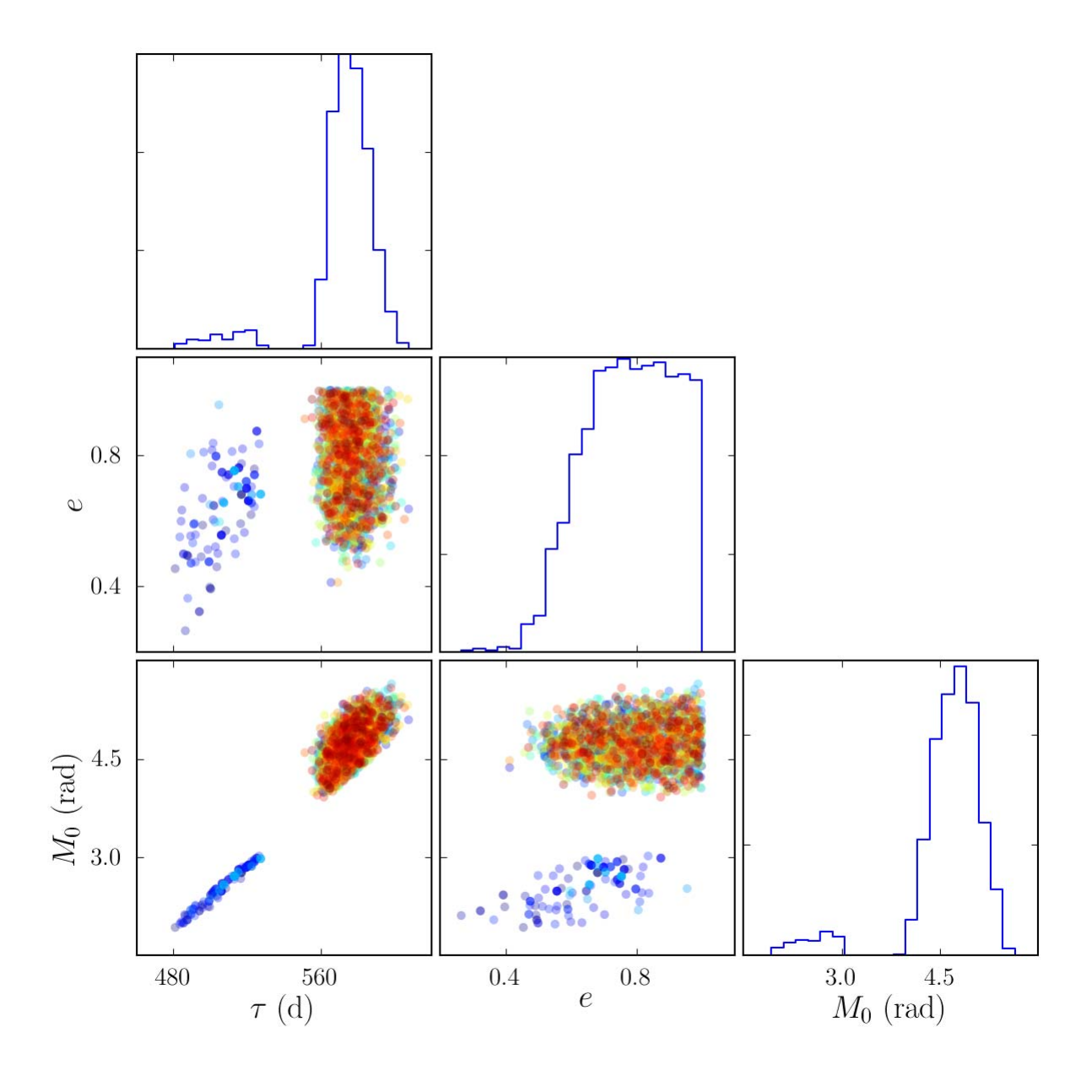}
\caption{\small Parameter estimation results from single-planet modeling of 24 RV observations
of HD~222582 reported in V00.  Shown are pairwise scatter plots and marginal histograms (along
the diagonal) for $\tau$, $\ecc$, $\mae$; color indicates time in Markov chain
(early to late in spectral sequence from blue to red).}
\label{fig:HD1b}
\end{figure}

Figures~\ref{fig:HD1a} and \ref{fig:HD1b} show selected results from this
approach, for a single-planet model for the 24 Keck RV observations from
HD~222582.  Figure~\ref{fig:HD1a} (top panel) shows the data as (red)
diamonds, at times earlier than $\sim 1400$~d (the $\approx$~3~m~s$^{-1}$
error bars are smaller than the symbols).  The dashed black curve shows the
best-fit orbit reported by V00, with a period of 579.5~d and a high
eccentricity, $e=0.71$. We used our pipeline to initialize and evolve a
population of 10 chains for 20,000 steps; convergence and mixing were
assessed via the Gelman-Rubin $R$ statistic (readily calculated from DEMC
parallel chains) and by inspection of trace plots and autocorrelation
functions.  The thin blue curves show the velocity curves for 20 models
chosen at random from the posterior samples (after thinning by a factor of
ten).  A wide variety of orbits are consistent with the data; no single
orbit fairly represents the possibilities (in fact, the best-fit orbit seems
not very representative). Figure~\ref{fig:HD1b} shows pairwise scatterplots
and marginal histograms (along the diagonal) of the posterior samples for
the nonlinear orbital parameters, summarizing the parameter uncertainties.
Sample points are transparently colored, indicating time in the Markov
chain; blue points are early in the chain, green and yellow points at
intermediate times, and orange and red points at late times.  For this
analysis the initial population was drawn from the marginal distribution
with $e=0$; its logarithm is essentially the LSP.  The separation of the
blue ``cool'' points from later points shows that the period for the peak of
the LSP was in the general vicinity of the global mode but did not
accurately locate it; all chains soon leave the vicinity of the initial
population (this burn-in phase of the chain would normally be discarded for
final inferences; we have kept it in this figure for illustration, but
discarded burn-in samples in later figures). The DEMC algorithm is able to
adapt to the posterior and find and explore the local mode in a few thousand
iterations (the burn-in time shortens to a few hundred steps with a more
sophisticated initialization using the full K-gram).

Posterior samples enable straightforward propagation of uncertainties into
nontrivial inferences.  For example, we can easily create samples from the
(marginal) posterior distribution for the physically important derived
parameters $a$ and $m_p\sin i$ (where $i$ is the inclination of the orbital
plane to the plane of the sky; $m_p$ and $i$ are not separately identifiable
in models for RV data) by evaluating them for each posterior sample. They
also enable us to tackle problems requiring prediction under uncertainty,
including adaptive scheduling of future observations (LC00, LC04, L04). 
From available data, $D$, we can calculate a posterior predictive
distribution for the value of a future datum at time $t$, $d_t$;
\be
p(d_t|D) = \int p(\params|D)\,p(d_t|\params)\, d\params,
\label{pdxn}
\ee
where $\params$ denotes all of the parameters in the currently
considered model.  The first factor in the integrand is the posterior
distribution for $\params$.  The second factor is the probability for
a future measured value, given values for the parameters; this is
just a normal distribution centered at the predicted velocity.  The
predictive distribution can be easily evaluated or sampled from using
posterior samples.

To optimally schedule a new observation, we must specify some measure of the
value a new observation would have for our observational goals; this is the
{\em utility function} in the decision-theoretical formulation of adaptive
scheduling, described in LC03.  For a general-purpose measure of the value
of new data for a variety of goals, we appeal to information theory:  we
seek new data that will make the updated posterior most informative about
the parameters, i.e., leading to the largest expected decrease in the
uncertainty, quantified by Shannon entropy (or, equivalently here, by
Kullback-Leibler divergence).  A straightforward calculation shows that the
posterior uncertainty is expected to decrease the most if we observe at the
time for which the {\em predictive} uncertainty---the entropy of
$p(d_t|D)$---is greatest: we learn the most by sampling where we know the
least.  This is called {\em maximum entropy sampling} (Sebastiani \& Wynn 2000).

We use posterior samples to calculate a straightforward Monte Carlo estimate
of the entropy in the predictive distribution as a function of time, $S(t)$.
The green curve in Figure~\ref{fig:HD1a} (expanded in the bottom panel)
shows $S(t)$ for times spanning about one orbit following the observations
reported in V00, in units of bits of information gain, relative to the
information that would be gained by repeating the last observation (one bit
of information roughly corresponds to reducing the volume of a joint
credible region by one half, a very significant information gain).  $S(t)$
essentially measures the vertical spread of the predicted velocity curves at
time $t$.  To reveal more detail in the predictive distribution as a
function of time, we show (magenta) point clouds of samples of $d_t$ from
$p(d_t|D)$ at six different times (the times are dithered over an interval
of $\pm 5$~d to spread the points).  The points reveal the predictive
distribution to be complicated, especially in regions of high uncertainty. 
It can be highly skew and even multimodal; clearly the predictive
uncertainties would not be well-described by normal distributions (which
underlie classical experimental design theory).  The entropy curve shows
that observations at the optimal time ($t\approx 1870$~d) are much more
informative than observations even a small fraction of a period earlier or
later.

\begin{figure}[t]
\centering
\includegraphics[width=\textwidth]{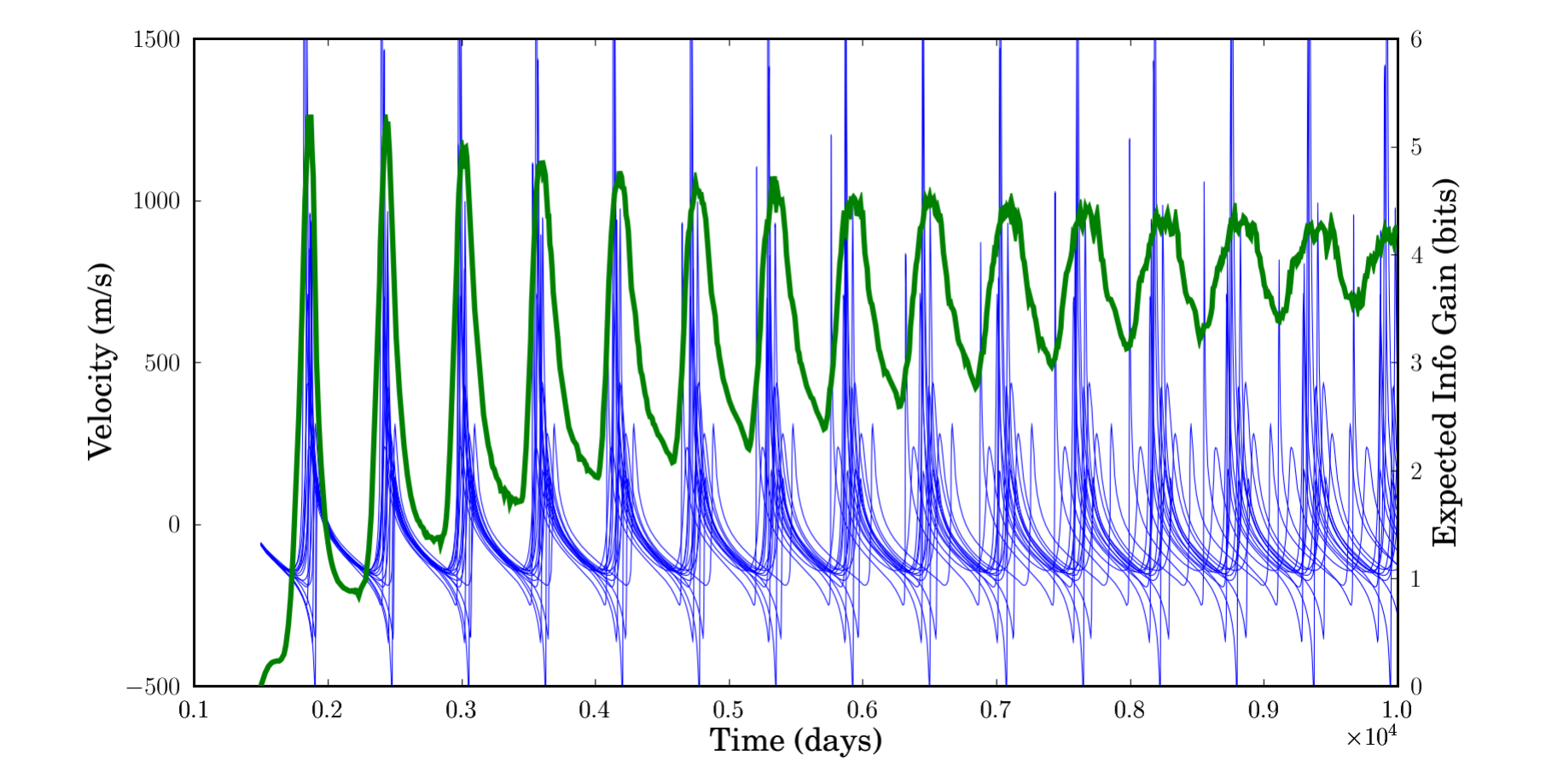}
\caption{\small Relative expected information gain for a single future
observation of HD~222582, for times $\approx 14$ orbits after original
observations (green curve, right axis), with 15 representative velocity models
(thin blue curves).}
\label{fig-HD2}
\end{figure}

Figure~\ref{fig-HD2} is a similar plot, but extending the $S(t)$ curve further
into the future, over about 14 orbits.  It shows that the most information
will be gained by following up the initial observations within the time
of the subsequent orbit or two.  There are definitely preferred times in
later orbits, but the maximum expected information gain is decreasing,
and the maximum and minimum of the expected information gain
are converging.  These features are in accord with intuition:  the
significant uncertainties in the orbital period and shape imply that
our ability to predict decreases with time, until after many orbits we
cannot even confidently predict the number of orbits that have passed.
At late times, our predictions become so poor that no candidate observing time
is clearly preferable to another.

We would like to see how adaptive scheduling can improve results compared to
non-optimal scheduling.  LC03 and L04 reported results of a few simulations
demonstrating the superiority of adaptive scheduling to informal or random
sampling.  Ford (2008) performed a more extensive simulation study, focusing
on the significantly simpler case of circular orbits, even more convincingly
demonstrating the benefits of the approach, though still with simulated data.

\begin{figure}[t]
\centering
\includegraphics[width=\textwidth]{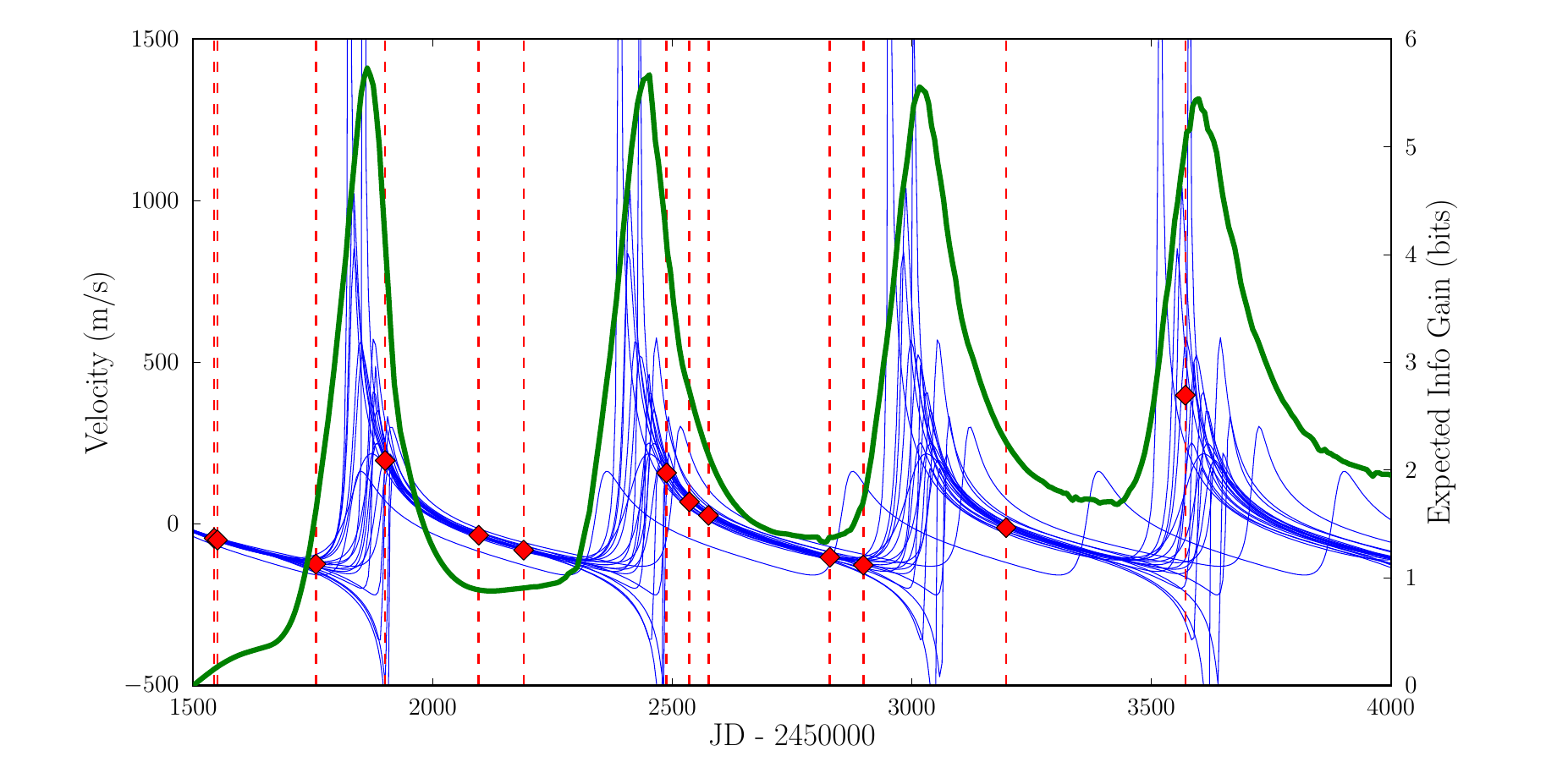}
\caption{\small Relative expected information gain for a single future
observation of HD~222582 for times $\approx 4$ orbits after
after the observations of V00 (thick green curve, right axis),
with 20 representative velocity models (thin blue curves, left axis).
Red diamonds indicated the the subsequent 13 observations of B06,
at times indicated by vertical dashed red lines.}
\label{fig:HD-next}
\end{figure}

A more convincing demonstration would use real data, for example, comparing
results from subsequent observing of HD~222582 with an optimized schedule to
results from a non-optimized schedule.  Butler et al.\ (2006; B06) reported 13
subsequent observations, at times not chosen to optimize information gain.
Figure~\ref{fig:HD-next} is similar to Figure~\ref{fig:HD1a}, but extending the
$S(t)$ curve farther in the future, over about 4 orbits, covering the span
of the new observations of B06.  While a few observations were not far from a
local optimum, many observations were at times offering much less expected
information gain than optimal times.  To see what might have been gained with
optimal observing, we simulate a new observation at the first (global) optimal
time, update interim inferences, calculate a new entropy curve, and repeat for
a few cycles.  We do not know the true orbit of HD~222582 to use for the
simulations; as a reasonable surrogate, we use the best-fit orbit reported by
B06 using all 37 observations.

\begin{figure}
\centering
\includegraphics[width=\textwidth]{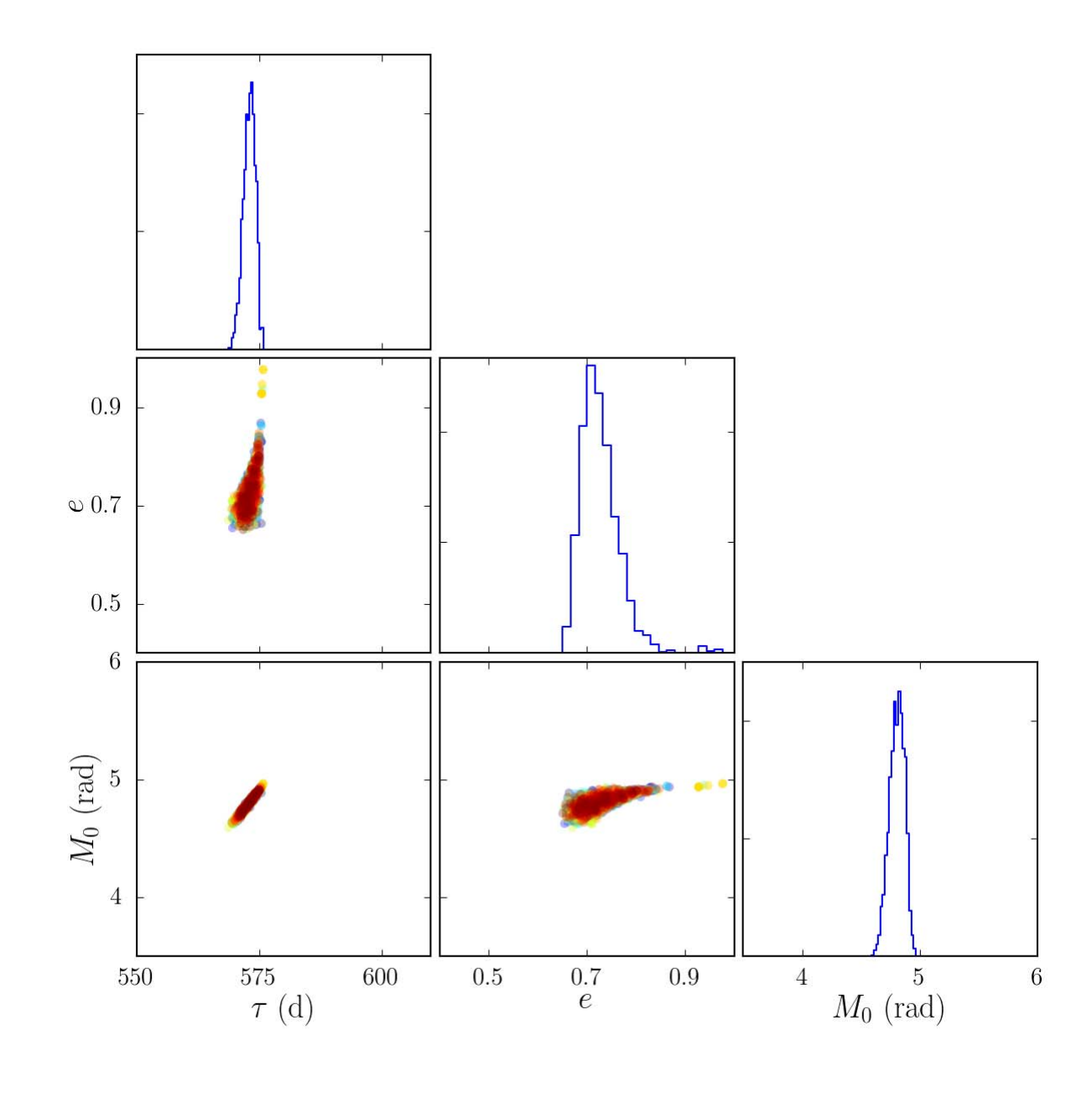}
\caption{\small Orbital parameter posterior samples for HD~222582 based on
24 early observations and three simulated new observations at optimal
times.}
\label{fig:HD-B06a}
\end{figure}

\begin{figure}
\centering
\includegraphics[width=\textwidth]{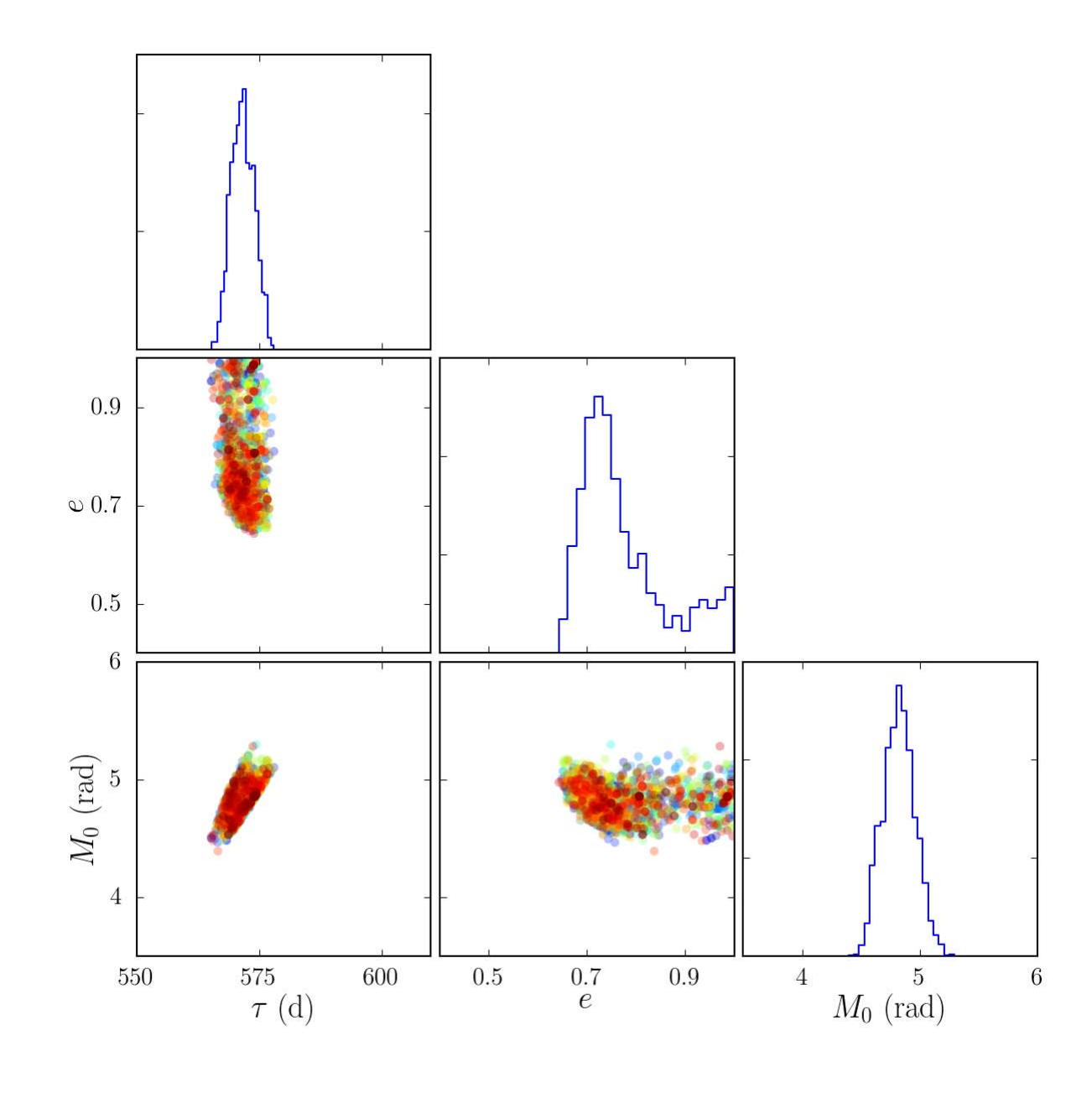}
\caption{\small Orbital parameter posterior samples for HD~222582 based on
24 early and 13 new, non-optimal actual observations (bottom).}
\label{fig:HD-B06b}
\end{figure}

Figures~\ref{fig:HD-B06a} and \ref{fig:HD-B06b} show posterior samples summarizing inferences with
new data.  Figure~\ref{fig:HD-B06a} shows results from using the 24 V00 observations and adding
just three new, optimal observations (for 27 total observations).  
Figure~\ref{fig:HD-B06b} shows results adding all 13 non-optimal observations from B06
instead of the three optimal observations (for 37
total observations).  The posterior distribution based on three new optimal
observations is dramatically more precise than the initial interim posterior
(displayed in Figure~\ref{fig:HD1b}), and is also much more precise than the
posterior based on 13 non-optimal observations (and it is converging on
the surrogate ``true'' parameter values).  This clearly demonstrates
the potential of adaptive observing for improving orbital parameter
estimates.  We are performing similar calculations for other heavily-observed
systems to more fully assess the benefits of the approach.


\section{Planet Detection Via Model Comparison}
\label{sec:detection}

To detect a planet, we must compare hypotheses with no planet (i.e, with
just $\vsys$ and $\sjit$ parameters), to hypotheses specifying a single
planet.  The comparison must account for the fact that both the no-planet
and single-planet models are {\em composite}; we do not know the best (or true)
parameter values a priori, and they remain uncertain (albeit less so)
even after fitting the best data sets.  To detect additional planets,
we must similarly compare multi-Kepler models to the single-planet model.
To optimally schedule observations for planet detection (as opposed to
orbit estimation, treated above), we must incorporate model uncertainty into 
prediction and experimental design calculations.

In Bayesian parameter estimation, the data influence inferences via the
likelihood function, the probability for the data considered as a function of
the parameters specifying a simple (point) hypothesis for the data. For
comparing rival parametric models, Bayes's theorem similarly indicates that
the data influence model choice through a likelihood---i.e., a probability for
the data given a hypothesis---but now the likelihood is for a hypothesis
specifying a model as a whole, i.e., a composite hypothesis.  This quantity may
fairly be called simply the likelihood for a model (as a whole); more
conventionally it is called the {\em marginal likelihood}, referring to how it
is calculated from the likelihood {\em function} for the model's parameters
(and helping to distinguish it linguistically from the likelihood function). 
The marginal likelihood is the integral of the product of a model's prior and
likelihood, over the entire parameter space.  Such integrals are often
challenging if the dimension of the parameter space is greater than a few.

For RV data, the no-planet model is only two-dimensional, and the mar\-gin\-al
likelihood integral over $(\vsys,\sjit)$ is easy to calculate.  Already
for the single-planet model, the parameter space is seven-dimensional and
the likelihood is highly structured; direct numerical cubature over all
seven dimensions is challenging.  If we are willing to adopt
interim priors allowing analytical amplitude marginalization, numerical
cubature is needed only over the three-dimensional nonlinear parameter space
(with a fourth dimension added when jitter uncertainty is important).  This
is tractable via cubature.  But adding a second or third planet makes 
marginal likelihood calculation via cubature intractable.  

The marginal likelihood needed for model comparison is just the normalization
constant for the posterior distribution used for parameter estimation.  We can
sample the posterior distributions for multi-planet models with the approach
of Section~\ref{sec:orbital}.  But the normalization constant does not
appear (and is not needed) in MCMC algorithms, so this ability does not
directly help us (although there are indirect ways to use MCMC output to
estimate marginal likelihoods; see Clyde et al.\ 2007).

We have developed a new method for estimating marginal likelihoods, based
on marrying ideas from adaptive importance sampling, mixture models, and
sequential Monte Carlo (SMC) methods.  At present we have implemented it
as a general-purpose ab-initio algorithm, without taking advantage of
any analytical marginalization or results from MCMC-based posterior
exploration.  Denote the full parameter space as $\params$, and let
$q(\params) = \pi(\params)\like(\params)$; the marginal likelihood is $Z =
\int q(\params) d\params$.  Suppose we could construct an importance sampling
density, $Q(\params)$, that resembles the target, $q(\params)$, but that
we could cheaply sample from and evaluate.  Then we could easily get accurate
$Z$ estimates via the usual importance sampling algorithm.  The problem is
that it is difficult to construct such importance samplers
(see, e.g., Oh \& Berger 1993, West 1993).

Instead of attempting to build an efficient importance density $Q(\params)$
ab initio, we build a {\em
sequence} of samplers, $Q_n(\params)$, approximating {\em annealed} versions
of the target, starting
with a nearly flattened target, and ending with the actual target.  Each
sampler is built using a mixture of multivariate Student-$t$ distributions
with $M$ components (setting the degrees of freedom to a small
value such as 5).  The number $M$, the component weights, and the locations
and scale matrices for each component are calculated using samples from the
previous step's sampler.  
In outline, we start with a dispersed importance density, $Q_0(\params)$,
and we set the initial annealed target density, $q_0(\params)$, equal
to $Q_0$.  The algorithm then cycles through the following steps
(starting with $n=1$) until
an acceptably small importance sampling variance is achieved:
\begin{enumerate}
\item Anneal the target from $q_{n-1}$ to 
$q_n = Q_0^{1-\beta_n} q^{\beta_n}$, where $\beta_n$ is a sequence
of ``inverse temperatures'' increasing from 0 to 1 according to an
annealing schedule (we use both standard rules of thumb for the schedule
and adaptive schedules; in our examples they worked equally well).
\item Sample parameter values $\{\params_i\}$ from $Q_{n-1}$; assign them
weights $w_i = q_n(\params_i)/Q_{n-1}(\params_i)$.
\item Refine $Q_{n-1}$ to $Q_n$, intended to approximate $q_n$:
  \begin{enumerate}
  \item Revise the Student-$t$ component parameters and mixture weights using
  the Expectation-Maximization (EM) algorithm to minimize the Kullback-Leibler
  divergence between $Q_n$ and $q_n$, estimated using the samples and sample
  weights.
  \item Delete mixture components with small mixture weights.
  \item Merge components that have large mutual information.
  \item Split components with large weights by duplicating them, revising their
  parameters via the EM algorithm, and keeping the split if the mutual information
  between the revised components is low.
  \end{enumerate}
\end{enumerate}
The final sampler can be used to estimate $Z$, and the
importance-weighted samples can be used for parameter estimation; the
algorithm thus may be able to replace our K-gram/DEMC parameter estimation
pipeline in certain settings.  We call the algorithm {\em annealing adaptive
importance sampling} (AAIS).  Liu et al.\ (2011; L11) provides a detailed
description; here we highlight some illustrative results.

\begin{figure}[t]
\centering
\centerline{\includegraphics[width=\textwidth]{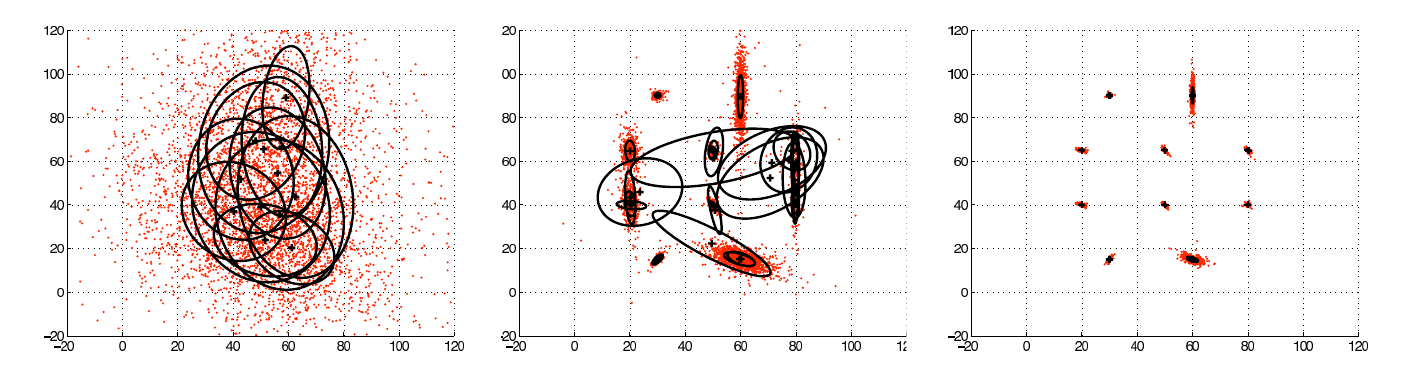}}
\caption{\small Three stages of the AAIS algorithm in operation on a
mixture of 10 well-separated bivariate normals with differing scales and
correlations.  From left to right, the annealed target has a temper
of $\beta_n = 0.01$, 0.11, and 1, corresponding to steps $n=1$, 3, and 8.}
\label{fig:AAIS-2D}
\end{figure}

Figure~\ref{fig:AAIS-2D} shows AAIS in operation on a highly multimodal
two-dimensional target with known marginal likelihood.  The target consists of
a mixture of 10 bivariate normals, all well-separated and with very different
covariance matrices.  
The initial importance sampler, $Q_0(\params)$, is a set of 10 bivariate $t$
distributions spread randomly across the space; it is illustrated by the
contours and crosses in the left panel. The annealing schedule has $\beta_n$
varying from 0.01 to 1 for $n=1$ to 8. We draw samples (red dots) from $Q_0$,
intended to approximate $q_0$; anneal the target to $q_1$; use $q_1$ to weight
the samples; and then use the weighted samples to adjust $Q_0$ to a new
sampler, $Q_1$, intended to approximate $q_1$, as outlined above. The middle
panel shows the third step ($\beta_3=0.11$), revealing that $Q_2$ is capturing
the features of the annealing target.  The right panel shows that in eight
steps all modes are located and well-modeled.  The final importance sampler,
$Q_8$, has an efficiency of 94\%, and estimates $Z$ to 0.25\% accuracy with
2000 samples (the accuracy estimate is from the usual importance sampling
variance).


\begin{figure}[t]
\centering
\centerline{\includegraphics[width=1.1\textwidth]{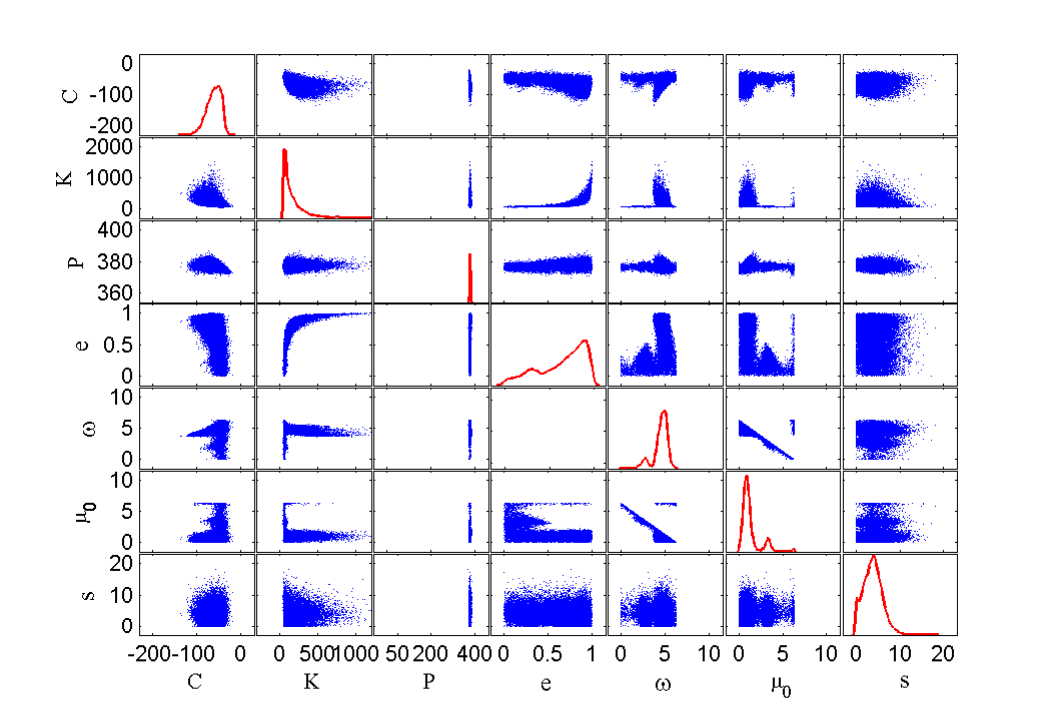}}
\caption{\small AAIS parameter estimates from an analysis of data from HD~73526;
shown are pairwise scatterplots and marginal histograms (curves, along the
diagonal) for the orbital parameters of the longer-period planet in a
two-planet model.  Renamed parameters (following L11) are:  $P=\oper$,
$C=\vsys$, $\mu_0=\mae$.}
\label{fig:AAIS-HD}
\end{figure}

Figure~\ref{fig:AAIS-HD} shows AAIS results from a two-planet fit to 30  RV
observations of HD~73526, a system known to contain two planets with orbital
periods of 188~d and 377~d (Tinney et al.\ 2006).  The panel shows parameter
estimates for the orbit of the second (longer-period) planet; the sampling
efficiency of the final sampler is $\approx 65$\%; similar or higher
efficiencies were obtained for single-planet and no-planet models.  The Bayes
factor (ratio of marginal likelihoods) for the single-planet model vs.\ the
no-planet model is $6.5\times 10^{6}$ ($\pm 3$\%); for two-planet vs.\
one-planet it is $8.2\times 10^{4}$ ($\pm 4$\%); this indicates very strong
evidence for two planets around HD~73526.  To provide confidence in the
estimates and uncertainties, we have validated AAIS by applying it to
a variety of multivariate test integrands more complicated than the 
illustrative two-dimensional case above, including integrands designed
to mimic key features of RV likelihood functions; L11 describes some
of these cases.

We are continuing to refine our K-gram/DEMC pipeline and the AAIS algorithm.
We are exploring how the algorithms compare for parameter estimation, and how
the K-gram may be used to accelerate the AAIS algorithm (via dimensional
reduction and a ``smart start'').  Our main longer-term goal is to generalize
the adaptive scheduling approach described in Section~3 for orbit estimation,
instead optimizing the timing of future observations for both planet detection
and orbit estimation simultaneously.  This requires incorporating AAIS in the
calculation (to handle planet number uncertainty), and considering non-greedy,
multiple-step scheduling designs.

{\em Acknowledgments}:  We are grateful to two referees whose comments helped
improve the manuscript, and to the editors for relaxing page constraints
to allow us to address the referees' questions.  Work reported here was
funded in part by NSF grants DMS-042240, AST-0507481, and AST-0507589, and by
the NASA {\em Space Interferometry Mission} via JPL subcontracts and the
{\em SIM} Science Studies program.




\newcommand{\jname}[1]{{\em #1}}
\newcommand{\aap}{\jname{Astron. \& Astrophys.}}
\newcommand{\aj}{\jname{Astronom. J.}}
\newcommand{\apj}{\jname{Astrophys. J.}}
\newcommand{\apjl}{\jname{Astrophys. J. Lett.}}
\newcommand{\mnras}{\jname{Mon. Not. Roy. Astron. Soc.}}
\newcommand{\pasp}{\jname{Pub. Astron. Soc. Pac.}}











\end{document}